\documentclass[12pt]{iopart}

\usepackage{graphicx}
\usepackage{cite}
\usepackage{floatflt}

\begin{document}

\title[Calculations of helium and hydrogen-helium plasma thermodynamics]
{Path integral Monte Carlo calculations of helium and
hydrogen-helium plasma thermodynamics and of the deuterium shock
Hugoniot}

\author{P R Levashov$^1$, V S Filinov$^1$, M Bonitz$^2$ and V E Fortov$^1$}

\address{$^1$Institute for High Energy Densities, RAS, Izhorskaya 13/19, Moscow 125412, Russia}
\ead{pasha@ihed.ras.ru}
\address{$^2$Christian-Albrechts-Universit\"at zu Kiel, Institut f\"ur
Theoretische Physik und Astrophysik, Leibnizstr. 15, 24098 Kiel, Germany}

\begin{abstract}
In this work we calculate the thermodynamic properties of
hydrogen-helium plasmas with different mass fractions of helium by
the direct path integral Monte Carlo method. To avoid unphysical
approximations we use the path integral representation of the
density matrix. We pay special attention to the region of weak
coupling and degeneracy and compare the results of simulation with
a model based on the chemical picture. Further with the help of
calculated deuterium isochors we compute the shock Hugoniot of
deuterium. We analyze our results in comparison with recent
experimental and calculated data on the deuterium Hugoniot.
\end{abstract}

\pacs{52.25.Kn, 52.27.Gr, 31.15.Kb, 62.50.+p}
\vspace{2pc}

\section{Introduction}

Hydrogen and helium are the most abundant elements in the
Universe, therefore thermodynamic properties of hydrogen and
helium plasmas are widely required for many astrophysical problems
\cite{Chabrier:Saumon:1992,Saumon:Chabrier:1995,Gudkova:Zharkov:1999:PSS,Nellis:2000:PSS}.
In particular, the investigation of the giant planets Jupiter and
Saturn, and to a lesser extent brown dwarfs demands the
thermodynamic information for hydrogen and helium in the
approximate range of temperatures $10^3 < T < 10^5$~K and mass
densities $0.01 < \rho < 100$~${\rm g/cm^3}$. This region is
characterized by coupling effects and chemical reactions caused by
partial pressure dissociation and ionization
\cite{Ebeling:BoundStates:1976,Kraeft:PACH1986}; these effects
considerably complicate an equation of state (EOS) calculation.
Moreover, in the same range of parameters the so-called plasma
phase transition (PPT) has been predicted by many authors
\cite{Norman:1968,
Ebeling:BoundStates:1976,Kraeft:PACH1986,Chabrier:Saumon:1992,EFFGP:Plasma:1991}.
However the application of the chemical picture
\cite{Ebeling:BoundStates:1976,Kraeft:PACH1986} at densities
corresponding to pressure ionization is questionable. Therefore
there is a great interest in direct first-principle numerical
simulations of strongly coupled degenerate systems which avoid
difficulties of conventional theories.

In this work we use the direct path integral Monte Carlo method
(DPIMC) to calculate thermodynamic properties of
hydrogen-helium plasma with different mass fractions of helium.
This method is well established
theoretically and allows the treatment of quantum and exchange
effects without any approximations using only fundamental physical constants. We
compare the results of our simulation with the EOS model based on
the chemical picture
\cite{Chabrier:Saumon:1992,Saumon:Chabrier:1995}. We also use the DPIMC method to compute the deuterium Hugoniot. We compare our simulation results with recent experimental and theoretical works and analyze the modern state of the problem.

\section{Simulation method and results for hydrogen-helium plasma}

The details of our computational scheme can be found elsewhere
\cite{Z-N-F:QMC1977,Filinov:2004:CPP,Filinov:2003:JPA:PPT,Filinov:2005:CPP}.
Modern supercomputers allow us to simulate about 100 quantum
particles in a Monte Carlo cell at a given temperature and volume.
The DPIMC \textit{has no limitations on coupling parameter} and can be
applied at \textit{significant degeneracy of the system} (with degeneracy
parameter values as high as 300) \cite{Filinov:2004:CPP}. Earlier the
method was thoroughly tested by simulating different properties of
ideal and interacting degenerate plasmas
\cite{Filinov:2000,Filinov:2000:PLA}. In particular, we
investigated temperature and pressure dissociation and ionization
{\it ab initio}; we also observed the effect of proton ordering at
very high densities and the formation of a Coulomb crystal of
protons \cite{Filinov:2000}.

In this section we calculate thermodynamic properties of
hydrogen-helium mixtures at relatively low coupling and degeneracy
parameters and compare our results with a well-known chemical
picture model used mostly in astrophysics
\cite{Saumon:Chabrier:1995, Chabrier:Saumon:1992}. This model
includes classical statistics for molecules and ions and
Fermi-Dirac statistics for the electrons. It takes into account
many physical effects including a number of subtle
"second-order" phenomena. We calculated thermodynamic properties
of hydrogen-helium mixtures with a composition corresponding to
that of the outer layers of the Jovian atmosphere. During the
mission of the Galileo spacecraft the helium abundance in the
atmosphere of Jupiter was determined as $Y =
{m_{\mathrm{He}}}/({m_{\mathrm{He}}+m_{\mathrm{H}}}) = 0.234$ and
was close to the present-day protosolar value $Y = 0.275$
\cite{Gudkova:Zharkov:1999:PSS}. As the model of the Jupiter is
significantly determined by its composition and EOS, it was
interesting to simulate the thermodynamic properties of the
mixture with different compositions in the region of pressure
dissociation and ionization.

\begin{figure}
  \includegraphics[width=6.5cm]{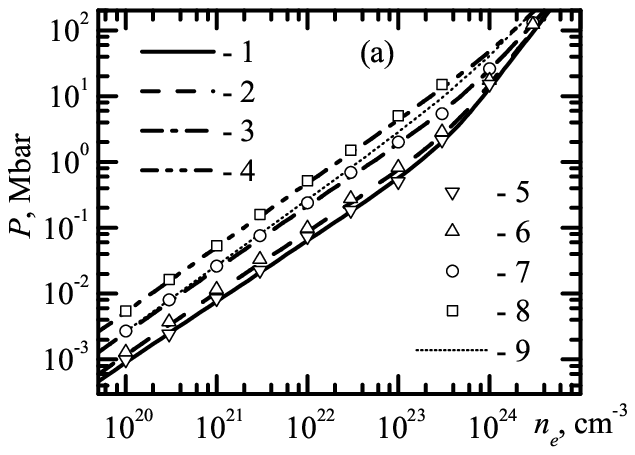}
  \includegraphics[width=6.5cm]{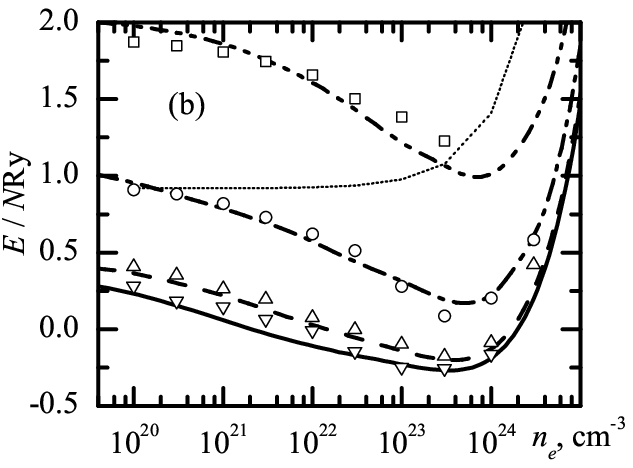}
  \caption{Pressure (a) and energy per particle (b) in a hydrogen-helium mixture with the mass concentration of helium Y = 0.234 ($\mathrm{Ry} \approx
13.6$~eV). Shown are DPIMC isotherms and related EOS isotherms
\cite{Saumon:Chabrier:1995}. EOS \cite{Saumon:Chabrier:1995}
(DPIMC) calculations: 1(5)~--— 40~kK, 2(6)~--- 50~kK, 3(7)~---
100~kK, 4(8)~--- 200~kK. 9~--— 100~kK isotherm for ideal plasma.}
\label{hhe40-200}
\end{figure}

We considered two mixtures with low and high abundance of helium.
The results of calculations for the mixture corresponding to the
outer layers of the Jovian atmosphere ($Y = 0.234$) in the region
of temperatures from $T = 10^4$ to $2\cdot 10^5$~K and electron
number densities from $n_e = 10^{20}$ to $3\cdot 10^{24}$~${\rm
cm^{-3}}$ are presented in \fref{hhe40-200}. The agreement between
our calculations and the model \cite{Saumon:Chabrier:1995} along
the isotherms $T = 4\cdot 10^4$, $5\cdot 10^4$, $10^5$, and
$2\cdot 10^5$~K is quite good and becomes better with the increase
of temperature. The formation of atoms and molecules is the reason
of the pressure and energy reduction along the $10^5$~K isotherm
with respect to the isotherm of a non-interacting hydrogen-helium
mixture (see \fref{hhe40-200}).

\begin{figure}
  \includegraphics[width=6.5cm]{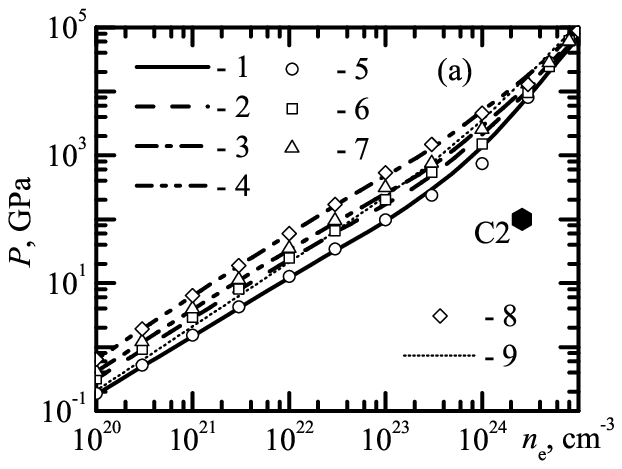}
  \includegraphics[width=6.5cm]{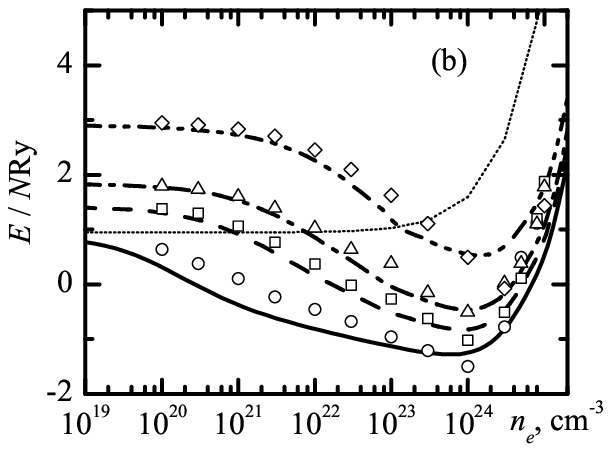}
  \caption{
Pressure (a) and energy per particle (b) in a hydrogen-helium mixture with
the mass concentration of helium $Y = 0.988$.  Shown are DPIMC isotherms and related EOS isotherms
\cite{Saumon:Chabrier:1995}. EOS \cite{Saumon:Chabrier:1995} (DPIMC)
calculations: 1(5) ~--- 100~kK, 2(6)~--- 156~kK, 3(7)~--- 200~kK,
4(8)~--- 312~kK.
9~--- 100~kK isotherm for ideal plasma. C2~--- critical point of the PPT \cite{EFFGP:Plasma:1991} ($T_{cr} \approx 120$~kK).
} \label{hhe_100_300}
\end{figure}

The results for $Y =
0.988$ (almost pure helium) at relatively high temperatures $T =
10^5\div 3\cdot 10^5$~K in a wide range of densities are presented in \fref{hhe_100_300}. The
agreement between our calculations and the model
\cite{Saumon:Chabrier:1995} along the isotherms $T = 10^5$,
$1.56\cdot 10^5$, $2\cdot 10^5$, and $3.12\cdot 10^5$~K is
satisfactory for pressure and internal energy per particle. The
smaller values of pressure on the DPIMC isotherms $10^5$ and
$1.56\cdot 10^5$~K near the particle density $10^{24}$~cm$^{-3}$
can be explained by a strong influence of interaction and bound
states in this region (see below). Ionization
effects also reduce the internal energy of the system in
comparison with non-interacting (ideal) plasma as it can be
clearly seen in Fig.~\ref{hhe_100_300}b. The positions of
ionization minima are well reproduced by the DPIMC method in a
good agreement with the chemical picture calculations. At higher
densities Fermi-repulsion gives the main contribution to pressure
and energy and this effect is also observed in our simulations.

At low temperatures $T < 3\cdot 10^4$~K and $Y = 0.234$ the agreement
between DPIMC and chemical picture calculations becomes worse, 
moreover, the region of thermodynamic instability has been discovered.
In particular, along the isotherm $T = 2\cdot 10^4$~K we have found 
such a region in the range
of densities between 0.5 and 5~g/cm$^3$. Along the isotherms $T =
1.5 \cdot 10^4$~K and $T = 10^4$~K this region is even wider and
begins from 0.38 g/cm$^3$ \cite{Filinov:2005:CPP}. Surprisingly, 
the region of DPIMC instability correlates 
with the range of temperatures ($T<2\cdot 10^4$~K) and densities (0.3--1~g/cm$^3$) 
in which the PPT in hydrogen or
hydrogen-helium mixture with low mass concentration has been predicted \cite{EFFGP:Plasma:1991,
Chabrier:Saumon:1992, Schlanges:1995}. Moreover, the sharp
rise of electrical conductivity of hydrogen-helium mixture along
the quasi-isentrope is also revealed experimentally in the range
of densities 0.5–-0.83~g/cm$^3$ \cite{Ternovoi:2004:HHe}. However, we cannot claim that 
these facts confirm the existence of PPT in our DPIMC simulation; in the nearest
future we plan to investigate the PPT problem in detail using more sophisticated numerical methods.  

Because of the high binding energy of electrons in He we currently
can obtain reliable results for $Y=0.988$ only at temperatures
higher than $10^5$~K. Under these conditions the influence of 
helium double ionization can lead to the formation of
bound states in the Monte Carlo cell as well as pressure
and internal energy decrease. Probably this effect takes place in
Fig.~\ref{hhe_100_300} near electron number density $n_e =
10^{24}$~cm$^{-3}$ at $T = 10^5$ and $1.56\cdot 10^5$~K; the
critical point of the possible PPT in this region with critical temperature $\approx
120$~kK \cite{EFFGP:Plasma:1991} is also shown in \fref{hhe_100_300}a.

\section{Deuterium shock Hugoniot}

\begin{figure}
  \begin{center}
    \includegraphics[width=10cm]{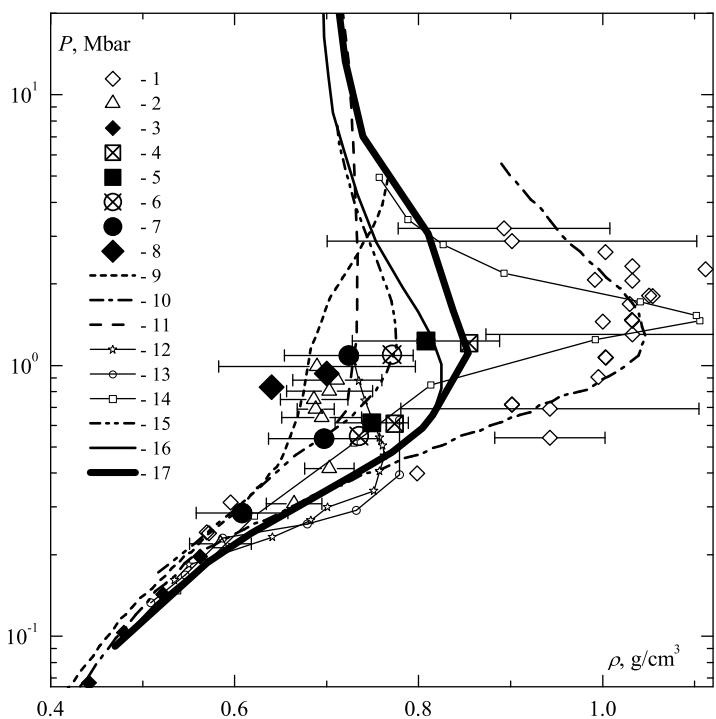}
  \end{center}
  \caption{Shock Hugoniot of deuterium. Experimental data for liquid deuterium: 1~--- \cite{DaSilva:1997, Collins:1998:Science},
2~--- \cite{Knudson:2004}, 3~--- \cite{Nellis:1983:JCP}, 6~--- \cite{Grishechkin:2004:JETPL}, and 7~--- \cite{Boriskov:2005:PRB}; for solid deuterium: 4~--- \cite{Belov:2002:JETPL, Boriskov:2003:DP}, and 5~--- \cite{Boriskov:2005:PRB}; for gaseous deuterium: 8~--- \cite{Grishechkin:2004:JETPL}. Calculations: 9~--- \cite{SESAME:1992},
10~--- \cite{Ross:1998:PRB}, 11~--- \cite{Militzer:2000}, 12~--- \cite{Desjarlais:2003:PRB}, 13~--- \cite{Bonev:2004:PRB}, 14~--- \cite{Knaup:2003}, 15~--- \cite{Grishechkin:2004:JETPL}, 16~--- \cite{Bezkrovniy:2004:PRE:1}, and 17~--- this study.}
  \label{dhug}
\end{figure}

Using our previous simulation results for deuterium we calculated the shock Hugoniot of liquid deuterium \cite{Filinov:2005:PPR}.
\Fref{dhug} summarizes the data from different
experimental, theoretical, and numerical studies on the
shock compression of deuterium. Measurements
performed in the NOVA facility, where a shock wave
in liquid deuterium with initial density 0.171 g/cm$^3$ was generated by a laser pulse \cite{DaSilva:1997, Collins:1998:Science} show that the deuterium
density behind the shock front can increase by a
factor of more than 6.
Experiments with the acceleration of an aluminum foil
by a magnetic field to velocities higher than 20~km/s
\cite{Knudson:2004} show a considerably lower compression ratio in
comparison to \cite{DaSilva:1997, Collins:1998:Science}. The results obtained in \cite{DaSilva:1997, Collins:1998:Science} and
\cite{Knudson:2004} disagree within experimental errors.

In contrast to
\cite{DaSilva:1997, Collins:1998:Science} and \cite{Knudson:2004}, where targets several hundred microns thick were used, in \cite{Belov:2002:JETPL, Trunin:Khariton03:D, Boriskov:2003:DP,Boriskov:2005:PRB}, the shock compressibility of solid (initial density 0.199~g/cm$^3$) \cite{Belov:2002:JETPL, Trunin:Khariton03:D, Boriskov:2003:DP} and liquid \cite{Boriskov:2003:DP} deuterium was measured in a 4-mm-thick layer
using a hemispherical explosive device. It is interesting to note that first such measurements for solid deuterium \cite{Belov:2002:JETPL, Boriskov:2003:DP} (points 4) showed greater compressibility of deuterium than it was reported later \cite{Boriskov:2005:PRB} (points 5 in \fref{dhug}). The same situation is observed for the experimental points on liquid deuterium (points 6 and 7, correspondingly, see \cite{Grishechkin:2004:JETPL} where preliminary experimental data for liquid deuterium from \cite{Boriskov:2005:PRB} are shown). Experimental points for liquid deuterium \cite{Boriskov:2005:PRB} are in a good correspondence with the data \cite{Knudson:2004}. Another hemispherical device was applied for shock loading of dense gaseous deuterium with initial density close to that of liquid deuterium \cite{Grishechkin:2004:JETPL}. In these experiments \cite{Grishechkin:2004:JETPL} apart from kinematic shock wave parameters temperature and light absorption of shock-compressed gas were registered. Two experimental points 8 corresponding to the initial gas densities 0.1335~g/cm$^3$ and 0.153~g/cm$^3$ are also shown in \fref{dhug}.
Curve 15 demonstrates the SAHA-IV liquid deuterium Hugoniot
with the initial density 0.171~g/cm$^3$ \cite{Grishechkin:2004:JETPL}.
The SAHA-IV chemical plasma model was calibrated so as to be in
agreement with points 8. In this case curve 15 passes through
the old position of the liquid Hugoniot point at 1.09~Mbar
\cite{Grishechkin:2004:JETPL}. The new position of the point
at 1.09~Mbar \cite{Boriskov:2005:PRB}, however, is shifted towards
 lower densities. Therefore points 7 and 8 in \fref{dhug} probably
 cannot be described by one and the same theoretical model.

In \fref{dhug} a number of calculated shock Hugoniots is also shown; the detailed analysis of these results can be found in our recent works \cite{Bezkrovniy:2004:PRE:2, Filinov:2005:PPR}. Here we can only indicate that the DPIMC Hugoniot is shifted towards higher densities in comparison to the experimental data published in
\cite{Knudson:2004, Boriskov:2005:PRB}. At pressures below 1–-2 Mbar, 
the thermodynamic instability revealed in \cite{Levashov:2001} comes
into play; therefore, a segment of the shock Hugoniot
that lies below 1~Mbar is not quite reliable.
At higher pressures the closest to the DPIMC Hugoniot is curve 16 calculated in \cite{Bezkrovniy:2004:PRE:1} by
the classical reactive ensemble Monte Carlo method. In
this method, the effects of dissociation of deuterium
molecules are taken into account most correctly; this
allows one to achieve good agreement with the experimental
data obtained at low temperatures and pressures
\cite{Nellis:1983:JCP}, even if ionization is not taken into account.
Therefore we combined the low-pressure part of the Hugoniot from
\cite{Bezkrovniy:2004:PRE:1} and high-pressure one from
\cite{Filinov:2005:PPR} at 15000~K and obtained the united Hugoniot
 \cite{Bezkrovniy:2004:PRE:2} (curve 17 in \fref{dhug}). We want to stress here that these two
methods are completely independent and no interpolation
procedure is used.

Thus we confirm that the experimental points \cite{DaSilva:1997,
Collins:1998:Science} are questionable and the true position of
the liquid deuterium Hugoniot remains unclear. We believe that
future experiments at the hemispherical device
\cite{Grishechkin:2004:JETPL} for densities of gaseous deuterium
corresponding to the liquid and solid states will give important
additional information about shock compression of liquid and solid
deuterium. In the nearest future we plan to calculate two DPIMC
Hugoniots corresponding to the initial gaseous deuterium densities
0.1335 and 0.153~g/cm$^3$ from the experiment
\cite{Grishechkin:2004:JETPL}.

\ack
This work is supported by the Deutsche Forschungsgemeinschaft via TRR 24,
the RAS program No.\ 17, the CRDF and the Ministry of
Education of Russian Federation Grants, and the RF President Grant No.~MK-3993.2005.8. 
The authors are also thankful to the Russian Science Support Foundation.

\section*{References}




\end{document}